\documentclass[a4paper,fleqn,usenatbib,letters]{mnras}

\usepackage{times}
\usepackage[T1]{fontenc}
\usepackage{ae,aecompl}

\usepackage{graphicx}	% Including figure files
\usepackage{amsmath}	% Advanced maths commands
\usepackage{amssymb}	% Extra maths symbols
\usepackage{color}
\usepackage{hyperref}
\title[Radio emission and jets in Her X-1]{Radio emission from the X-ray pulsar Her X-1: a jet\\
launched by a strong magnetic field neutron star?}

\author[Van den Eijnden et al.]{
\noindent J. van den Eijnden$^{1}$\thanks{E-mail: a.j.vandeneijnden@uva.nl},
N. Degenaar$^{1}$,
T. D. Russell$^{1}$,
J. C. A. Miller-Jones$^{2}$,
\newauthor R. Wijnands$^{1}$,
J. M. Miller$^{3}$,
A. L. King$^{4}$ and
M. P. Rupen$^{5}$\\
$^{1}$Anton Pannekoek Institute for Astronomy, University of Amsterdam, Science Park 904, 1098 XH Amsterdam, The Netherlands\\
$^{2}$International Centre for Radio Astronomy Research -- Curtin University, GPO Box U1987, Perth, WA 6845, Australia\\
$^{3}$Department of Astronomy, University of Michigan, 500 Church Street, Ann Arbor, MI 48109, USA\\
$^{4}$KIPAC, Stanford University, 452 Lomita Mall, Stanford, CA 94305, USA\\
$^{5}$Herzberg Astronomy and Astrophysics Research Centre, 717 White Lake Road, Penticton, BC, V2A 6J9, Canada}

\date{Accepted XXX. Received YYY; in original form ZZZ}

\pubyear{2017}

\begin{document}
\label{firstpage}
\pagerange{\pageref{firstpage}--\pageref{lastpage}}
\maketitle

\begin{abstract}
\noindent Her X-1 is an accreting neutron star in an intermediate-mass X-ray binary. Like low-mass X-ray binaries (LMXBs), it accretes via Roche-lobe overflow, but similar to many high-mass X-ray binaries containing a neutron star, Her X-1 has a strong magnetic field and slow spin. Here, we present the discovery of radio emission from Her X-1 with the Very Large Array. During the radio observation, the central X-ray source was partially obscured by a warped disk. We measure a radio flux density of $38.7 \pm 4.8$ $\mu$Jy at $9$ GHz but can not constrain the spectral shape. We discuss possible origins of the radio emission, and conclude that coherent emission, a stellar wind, shocks and a propeller outflow are all unlikely explanations. A jet, as seen in LMXBs, is consistent with the observed radio properties. We consider the implications of the presence of a jet in Her X-1 on jet formation mechanisms and on the launcing of jets  by neutron stars with strong magnetic fields. 
\end{abstract}


%%%%%%%%%%%%%%%%%%%%%%%%%%%%%%%%%%%%%%%%%%%%%%%%%%

%%%%%%%%%%%%%%%%% BODY OF PAPER %%%%%%%%%%%%%%%%%%

\section{Introduction}

Her X-1 is an extensively-studied accreting X-ray pulsar, discovered with the \textit{UHURU} satellite \citep{tananbaum72}. The pulsar has a low spin period of $1.24$ s and is in a binary system with an orbital period of 1.7 days \citep{leahy14}. Her X-1 was the first accreting neutron star (NS) where a cyclotron line was discovered \citep{trumper78}, with an energy of $\sim 40$ keV. Although this energy varies with time and X-ray flux \citep[e.g.][]{staubert16}, it provides a direct measurement of the pulsar magnetic field of a few times $10^{12}$ G. 

Her X-1 shows peculiar variability in X-rays over a $35$-day cycle, originating from the precession of a warped accretion disk \citep[][see also Fig. \ref{fig:lc}]{scott99}: the cycle starts in the bright Main High (MH) state, offering an unobscured view of the central X-ray source. This is followed with the Low State (LS), where the X-ray flux drops $\sim 99\%$ and only reflection off the face of the companion and an accretion disk corona are visible \citep{abdallah15}. This LS is interspersed by the Short High (SH) state, reaching a few tens of percent of the original MH state flux. This variability is geometric, and the central X-ray source does not intrinsically vary on the $35$-day timescale. 

The companion star of Her X-1 has a mass of $2.2M_{\odot}$ \citep{reynolds97,leahy14}. At the simplest level, accreting NSs are classified based on the donor star mass into low-mass X-ray binaries (LMXBs, $\lesssim 1 M_{\odot}$), high-mass X-ray binaries (HMXBs, $\gtrsim 10 M_{\odot}$) and the rare intermediate-mass X-ray binaries (IMXBs) inbetween. Her X-1 is an IMXB but combines characteristics from both other classes: as in LMXBs, it accretes through Roche-lobe overflow and an accretion disk \citep{scott99}, while most HMXBs accrete from the wind or circumstellar disk of the donor. In addition, Her X-1 has the strong magnetic field and low spin that are typically seen in HMXBs; NS LMXBs instead tend to have weaker magnetic fields of $B \lesssim 10^9$ G and if pulsations are seen, these are typically at millisecond periods \citep{patruno12}. 

Another observational difference between HMXBs and LMXBs is the presence of radio emission and inferred jets. LMXBs very commonly show synchrotron emission from jets, which is correlated with the X-ray emission from the accretion flow \citep[][]{migliari06,tudor17,gusinskaia17}, similar to accreting black holes \citep[BHs;][]{merloni03,falcke04}. On the contrary, in NS HMXBs jets have only been observed in Cir X-1, a young NS that might have a high-mass donor \citep{johnston16}. As jet formation is still poorly understood, it is unclear which properties of NS LMXBs and HMXBs could explain this apparent systematic difference: the spin period, magnetic field, or the presence of an accretion disk might all play a vital role. As Her X-1 shares characteristics of both classes, it can help understand the difference between their jet launching abilities.  

In this Letter, we present the discovery of radio emission from Her X-1. We present the observations and results in Sections 2 and 3, and afterwards discuss the origin of the emission and implications for our understanding of jet formation in LMXBs and HMXBs. We also consider the possibilities for future observations. 

\section{Observations}

\subsection{Radio}

We observed Her X-1 with the Karl G. Jansky Very Large Array (VLA) on 06 June 2013 (MJD 56449) from 02:12:01 to 03:26:38 UT, for a total of $\sim 54$ minutes of on-source observing time (project ID: 13A-352, PI: Degenaar). We observed the target in X-band between $8$ and $10$ GHz in two basebands, while the array was in C-configuration, yielding a synthesized beam of $3.24"\times1.8"$ (position angle $8.57^{\rm o}$). We used J1331+305 and J1635+3808 ($5.3^{\rm o}$ from the target) as the primary and secondary calibrators, respectively.   

The observation was calibrated and imaged following standard procedures with the Common Astronomy Software Applications package (CASA) v4.7.2 \citep{mcmullin07}. We did not encounter any significant RFI or calibration issues. Using CASA's multi-frequency, multi-scale \textsc{clean} task, we imaged Stokes I and V to make a source model of the field. With Briggs weighting and setting the robustness parameter to $0$ to balance sensitivity and resolution, we reached an RMS noise of $4.8$ $\mu$Jy\,beam$^{-1}$. We fit a point source in the image plane by forcing the fit of an elliptical gaussian with the FWHM and orientation of the synthesized beam. In addition, we also individually imaged the $8$--$9$ and $9$--$10$ GHz basebands with the same approach as the full band. As we did not observe a polarization calibrator, beam squint can affect our circular polarization estimates away from the pointing centre by a few percent.

\subsection{X-rays}

\begin{figure}
  \begin{center}
    \includegraphics[width=\columnwidth]{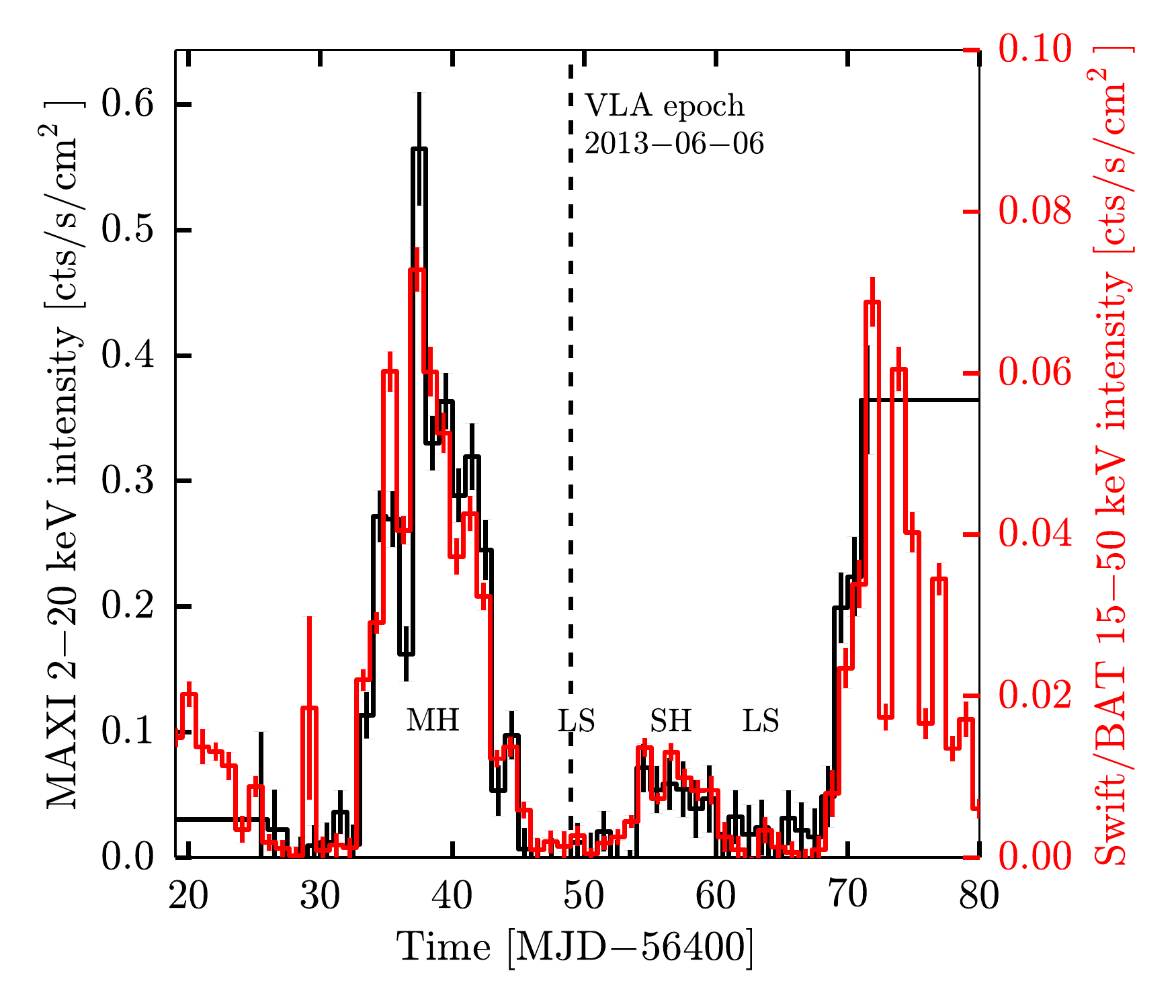}
    \caption{\textit{MAXI}/GSC and \textit{Swift}/BAT daily X-ray lightcurves of Her X-1 around the radio epoch on MJD 56449. The $35$-day cyclic variability, due to precession of the warped accretion disk, is clearly visible. The VLA epoch is shown by the dashed line.}
    \label{fig:lc}
  \end{center}
\end{figure}

We examined the X-ray properties of Her X-1 during the VLA epoch in order to obtain a simultaneous X-ray flux and determine the source's phase in the $35$-day precession cycle. To measure the X-ray flux, we extracted the \textit{MAXI}/Gas Slit Camera \citep[GSC;][]{matsuoka09} spectrum for the MJD of the VLA observation from the \textit{MAXI} website (\href{http://maxi.riken.jp}{http://maxi.riken.jp}). We extracted the spectrum for the full MJD to ensure a sufficient number of counts for a basic characterisation of the spectrum. We also obtained the \textit{MAXI}/GSC and \textit{Swift}/Burst Alert Telescope (BAT) \citep{krimm13} long term X-ray lightcurves of Her X-1. Fig. \ref{fig:lc} shows the \textit{MAXI} and \textit{Swift} light curves, clearly showing the $35$-day cycle and revealing that Her X-1 was in the first low-state of its precession cycle. Finally, we also downloaded the \textit{MAXI} spectrum on MJD 56437, the peak of the prior MH state, to estimate the unobscured X-ray flux. 

\section{Results}

\subsection{Radio}

\begin{figure}
  \begin{center}
    \includegraphics[width=\columnwidth]{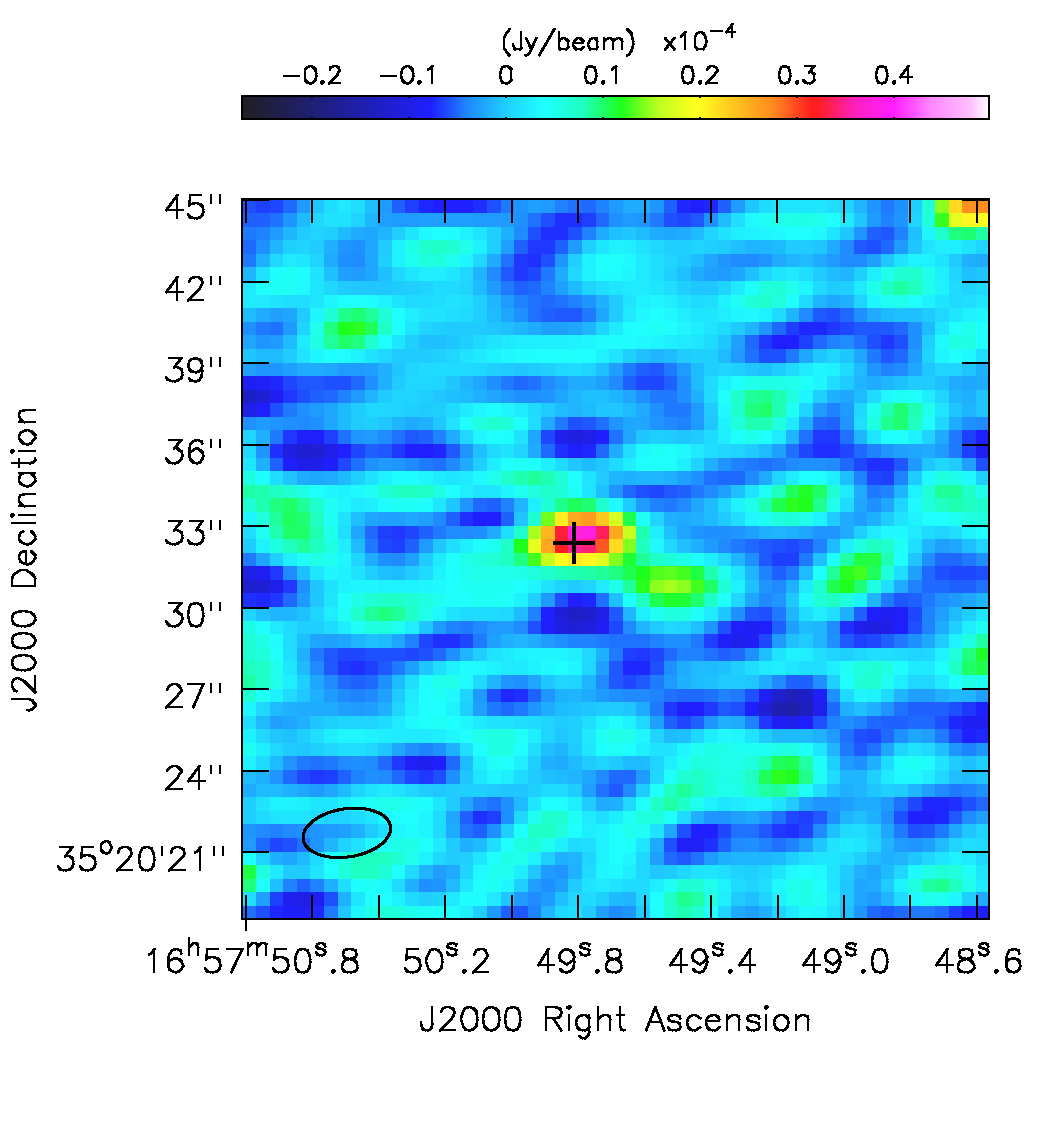}
    \caption{VLA image of Her X-1 at $9$ GHz. The black cross indicates the best known position, from the infrared 2MASS survey. In the bottom left corner, we show the half-power contour of the synthesized beam}
    \label{fig:image}
  \end{center}
\end{figure}

Her X-1 is detected at a flux density $S_{\nu} = 38.7 \pm 4.8$ $\mu$Jy at $9$ GHz, with a significance of $8\sigma$. A zoom of the target field is shown in Fig. \ref{fig:image}. For a distance of $d = 6.1$ kpc \citep[e.g][]{leahy14} and defining the radio luminosity $L_R = 4\pi\nu S_{\nu}d^2$, this corresponds to $L_R = (1.6 \pm 0.2)\times10^{28}$ erg s$^{-1}$. The source is also detected in the $8$--$9$ and $9$--$10$ GHz bands separately at $42.2 \pm 6.8$ and $36.2 \pm 6.8$ $\mu$Jy respectively. However, the low significance means we do not well constrain the radio spectrum with $\alpha = -0.7 \pm 5.3$, where $S_{\nu} \propto \nu^{\alpha}$. 

We measured a position of $\rm RA = 16^{\rm h}57^{\rm m}49^{\rm s}.792\pm0^{\rm s}.027$ and $\rm Dec = +35^{\rm o}20'32".578 \pm 0".225$, where the uncertainties equal the synthesized beam size divided by the signal-to-noise of the detection. This position is consistent within the $1-\sigma$ errors with the best known position of Her X-1, from the infrared 2MASS survey \citep{skrutskie06}, which is shown in Fig. \ref{fig:image} as well, and with the lower-accuracy positions at other wavelenghts. Hence, this is unlikely to be a background source. 

\subsection{X-rays}

We fit the two downloaded \textit{MAXI} $2$--$20$ keV spectra of Her X-1 to determine the flux on the MJD of the radio observation and at the height of the previous MH state. As the latter is only a short ($\sim 120$s) exposure, both spectra contain few photons ($\sim 145$ and $40$ photons, respectively) and are only suitable for a very simple fit. In both cases, we used \textsc{xspec} to fit an absorbed (\textsc{tbabs}) blackbody (\textsc{bbodyrad}) spectrum. We fix the $N_H$ in both cases, as the data quality is not sufficient to determine it directly. We set $N_H = 1.0\times10^{22}$ for the obscured LS \citep{inam05} and $N_H = 1.7\times10^{20}$ cm$^{-2}$ in the HS \citep{furst13}. This yields $0.5$--$10$ keV X-ray fluxes of $\sim 9\times10^{-11}$ erg s$^{-1}$ cm$^{-2}$ during the radio observation and $\sim 3\times10^{-9}$ erg s$^{-1}$ cm$^{-2}$ during the MH state. The latter is slightly lower than the typical range of X-ray fluxes of Her X-1 in the MH state of $5\times10^{-9}$ to $10^{-8}$ erg s$^{-1}$ cm$^{-2}$ \citep{staubert16}. This difference might be due to the short exposure of the \textit{MAXI} spectrum, combined with the dips often seen during the MH state \citep{igna11}. 

\section{Discussion}
\label{sec:discussion}

We present the first radio detection of the IMXB Her X-1, at a flux density of $S_{\nu} = 38.7 \pm 4.8$ $\mu$Jy. Her X-1 has been the subject of multiple radio searches, but similar to most NS HMXBs, was hitherto never detected. \citet{coe80} observed Her X-1 every day of an entire $35$-day precession cycle. The source was not detected in any of the observations, with $3\sigma$ upper limits of $9$ mJy. \citet{nelson88} observed Her X-1 twice in a large sample study of X-ray binaries and cataclysmic variables, reaching a $5\sigma$ upper limit of $1.3$ mJy. In this discussion, we will first compare the radio properties of Her X-1 with different classes of accreting NSs. Subsequently, we will discuss the origin of the radio emission and implication for future research. 

\subsection{Comparison with NS LMXBs and HMXBs}

Radio detections and jets are ubiquitous in disk-accreting, weak magnetic field NS LMXBs \citep[][]{migliari06,tudor17,gusinskaia17}. While $L_X$ and $L_R$ do appear to be related for these types of NS systems, no universal relation has emerged \citep{tudor17}. Most relevant for the comparison with Her X-1 are the handful of LMXBs containing a slow pulsar. With the exception of one (see below), none of these sources have been detected in the radio. 2A 1822-371 and 4U 1626-67 have unconstraining upper limits on their radio flux of $200$ $\mu$Jy \citep{fender00}. GRO 1744-28 (The Bursting Pulsar) does have deep \textit{ATCA} upper limits during it small 2017 outburst \citep{russell17}. Finally, for the mildly recycled $11$-Hz pulsar IGR J1748-2466 no radio upper limits are known. 

As stated, a single slow pulsar in an LMXB has been detected: the symbiotic X-ray binary GX 1+4 was recently discovered in radio ({\color{blue} Van den Eijnden et al., submitted}). In this type of sources, the NS accretes from the stellar wind of an evolved low-mass companion. The origin of the radio emission in GX 1+4 can not be unambigiously inferred. Other symbiotic X-ray binaries have not been targeted by radio campaigns and it is thus unknown whether radio emission occurs is more sources of this type. Given the current upper limits or lack of observations, new, deep radio observations are needed to infer whether Her X-1 is an outlier among slow pulsars in LMXBs or whether radio emission occurs more commonly among such sources. 

Her X-1 also shares characterics with many of the NS HMXBs: a strong magnetic field ($\gtrsim 10^{12}$ G) and a slow spin. Radio detections of HMXBs are relatively rare \citep{duldig79,nelson88,fender00,migliari11b}: the NS Cir X-1 launches resolved radio jets \citep{tudose06}, and likely is a HMXB (see \citealt{johnston16} for a recent discussion). However, no magnetic field estimate is known for Cir X-1. Two other NS HMXBs have been detected in radio. Most notably, the wind-accreting HMXB GX 301-2 was detected over multiple radio epochs \citep{pestalozzi09}. However, the flux levels were consistent with those expected from the stellar wind, and the claimed transient outflow component in the emission has not been confirmed \citep{migliari11b}. Additionally, the Be/X-ray binary A 1118-61, consisting of a NS and a Be companion, was detected in only one out of eight observations by \citet{duldig79}. Due to the crowded field, this detection might not be related to the Be/X-ray binary. Both these (possible) detections are thus not conclusive about the presence of a jet. 

There exist numerous radio non-detections of NS HMXBs. However, for most of these sources, the radio upper limits \citep[ranging from hundreds of $\mu$Jy to mJy levels;][]{duldig79,nelson88,fender00} are not constraining compared with NS LMXBs and deep observations with current generation radio telescopes might reveal these sources. Only the Be/X-ray binaries A 0535+26 \citep{tudose10} and X Per, and the wind-accreting NS HMXB 4U 2206+54 \citep{migliari06} have radio luminosity upper limits of $\lesssim 5\times10^{27}$ erg/s, below our Her X-1 measurement. However, these sources were observed at much lower (more than an order of magnitude) X-ray luminosity, making any direct comparison with Her X-1 difficult. 

\subsection{The emission mechanism and physical origin}

Three radio emission mechanisms are relatively unlikely to explain our detection of Her X-1. Firstly, thermal emission would require too high densities of emitting material on too large scales. Secondly, we imaged Stokes V in addition to Stokes I and did not detect the target, setting a $3\sigma$ upper limit on the circular polarization of $37\%$. Coherent emission should be highly circularly polarized and can thus be excluded. Finally, free-free emission from a strong stellar wind, as observed in the HMXB GX 301-2 \citep{pestalozzi09}, is unlikely: while a wind might be present \citep{leahy15}, its strength implies a flux density over two orders of magnitude lower than our detection \citep{wright75}. On the constrary, synchrotron emission is consistent with the observed radio properties. In the following, we will discuss possible physical origins of such synchrotron emission.

Firstly, synchrotron-emitting shocks could occur in the interaction between the disk and the magnetosphere or in the accretion column onto the magnetic poles. However, the Compton limit on the brightness temperature of $10^{12}$ K sets a lower limit on the size of the emitting region of $\gtrsim 7.5\times10^4$ km. We can estimate the size of the magnetosphere $R_{\rm m}$ by rewriting equation 1 from \citet{cackett09}:
\begin{equation}
\begin{split}
R_{\rm m} = k& \left(\frac{B}{1.2\times10^5 \rm G}\right)^{4/7} \left(\frac{f_{\rm ang}}{\eta}\frac{F_{\rm bol}}{10^{-9} \rm erg~s^{-1}~cm^{-2}}\right)^{-4/14} \\
&\left(\frac{M}{1.4M_{\odot}}\right)^{-8/7} \left(\frac{R}{10\rm km}\right)^{-12/7} \left(\frac{D}{5 \rm kpc}\right)^{-4/7} \text{ } R_g
\end{split}
\end{equation}
where $k$ is a geometry factor relating spherical and disk accretion, typically assumed to be $0.5$ for disk accretion, $B$ is the magnetic field strength, $f_{\rm ang}$ is the anisotropy correction factor, $\eta$ is the accretion efficiency, $F_{\rm bol}$ is the bolometric flux, and $M$, $R$ and $D$ are the mass, radius and distance of the NS. We use $B \sim 3\times10^{12}$ G \citep{staubert16}, $k=0.5$, $f_{\rm ang}=1$, $\eta=0.1$, $M=1.4M_{\odot}$, $R=10$ km \citep{leahy04} and D=$6.1$ kpc \citep{leahy14}. 

As $R_m$ scales inversely with flux, the maximum magnetospheric size can be estimated with the LS $2$--$10$ keV X-ray flux without a bolometric correction (e.g. $9\times10^{-11}$ erg s$^{-1}$ cm$^{-2}$): this yields $R_m \approx 1.7\times10^4$ km, smaller than the minimum emission region size. As the low flux during the LS of Her X-1 originates from a geometric effect, it is actually more accurate to use the MH state bolometric flux; for the measured MH state flux of $3\times10^{-9}$ erg s$^{-1}$ cm$^{-2}$, $R_m$ is even smaller at $\sim 0.7\times10^4$ km. Hence, shocks can be exluded as well, assuming that they indeed occur at the magnetosphere and not further out in the accretion flow. 

Another possibility is that we observe a propeller-driven outflow: if the magnetosphere spins faster than the disk where the magnetic and gas pressure are equal, it creates a centrifugal barrier that can either trap the disk \citep{dangelo10} or expel the material \citep{illarionov75,campana98}. The latter has, for instance, recently been inferred through X-ray monitoring in several strong magnetic field accreting NSs \citep[e.g. two Be/X-ray binaries;][]{tsygankov16b} and might explain the recent radio detection of GX 1+4 ({\color{blue} Van den Eijnden et al., submitted}). For a given NS magnetic field and spin period, one can estimate the maximum $L_X$ for which the magnetosphere can still create this centrifugal barrier as \citep[e.g.][]{campana02}:
\begin{equation}
\begin{split}
L_{X,\rm max} \approx 4\times10^{37} &k^{7/2} \left(\frac{B}{10^{12} \rm G}\right)^{2} \left(\frac{P}{1\rm s}\right)^{-7/3} \\ 
&\left(\frac{M}{1.4M_{\odot}}\right)^{-2/3} \left(\frac{R}{10 \rm km}\right)^{5} \text{ } \rm erg~s^{-1}
\end{split}
\label{eq:prop}
\end{equation}
where $P$ is the pulsar spin and all other parameters are already defined. For a magnetic field of $\sim 3\times 10^{12}$ G, a spin period of $1.24$ s and standard NS parameters, we estimate the $L_{X,\rm max} \approx 1.9\times10^{37}$ erg s$^{-1}$ for Her X-1. 

To assess whether a magnetic propeller could be at play in Her X-1, we need to compare this maximum X-ray luminosity with the correct $L_X$ of Her X-1. During the LS radio epoch, $L_X \approx 4\times10^{35}$ erg s$^{-1}$, comfortably below the upper limit for the propeller effect. However, the actual, unobscured X-ray luminosity is the more accurate probe of the relevant physical properties (i.e. the mass accretion rate that balances the magnetic pressure). During the prior MH state, Her X-1 reached $L_X \approx 1.2\times10^{37}$ erg s$^{-1}$ between $2$--$10$ keV. With a bolometric correction, the X-ray luminosity of Her X-1's MH state typically reaches $(2.5-5)\times10^{37}$ erg s$^{-1}$ \citep{staubert16}. This is of comparable magnitude as the $L_{X,\rm max}$ estimate, although it should be noted that not every MH state reaches the same luminosity \citep[e.g.][]{staubert16} and Eq. \ref{eq:prop} is merely an estimate. However, in other accreting NSs propellers have been linked to a simultaneous decrease in X-ray flux and pulsation strength and such behaviour has not been observed in Her X-1 in its regular, $35$-day cyclic behaviour. 

Finally, we might observe a compact, synchrotron-emitting radio jet, similar to those seen in NS LMXBs with weaker ($\lesssim 10^9$ G) magnetic fields. If we compare Her X-1's radio properties with the NS LMXB sample in the $L_X$/$L_R$ diagram \citep[see e.g.][for recent versions]{tudor17,gusinskaia17}, we see that these are consistent with several AMXP observations if we assume the LS flux. While an interesting comparison, as AMXPs have $\sim3-4$ orders of magnitude weaker magnetic fields and faster spins, we should actually again use the estimated unobscured (i.e. MH state) flux. In that comparison, the $L_R$ of Her X-1 is three to ten times lower than hard-state and several soft-state Atoll sources at similar $L_X$, and more similar to that of jet-quenched sources \citep[e.g.][]{gusinskaia17}.

As jet formation is poorly understood, the cause of jet quenching is puzzling. It is observed in all BH LMXBs if and when they transition into the soft spectral state \citep[e.g.][]{gallo03}, but the picture is more ambigious in accreting NSs: only in a handful of sources is quenching observed \citep[see e.g.][]{millerjones10, gusinskaia17}. If jet formation requires large scale-height poloidal magnetic fields, quenching might be explained as follows: in the hard spectral state, the accretion disk might be truncated away from the compact object \citep[e.g.][]{done07} as a radiatively-inefficient accretion flow \citep[RIAF;][]{narayan95} or corona replaces the inner disk, providing the required fields. As the disk moves inwards during the transition to softer spectral states, the RIAF disappears or the corona is cooled, breaking the jet formation mechanism. 

If the above scenario indeed underlies jet quenching, it is difficult to reconcile with the low $L_R$ in Her X-1: there, the strong stellar magnetic field prevents the disk from moving inwards. However, this stellar field might instead hamper the initial formation of a RIAF or corona, also effectively quenching the jet. Alternatively, the jet formation might be partially suppressed as the disk pressure can not dominate and twist the strong magnetic field \citep{massi08}, or as the magnetic field prevents the formation of a boundary layer at the NS surface, which might play a role in NS jet formation \citep{livio99}.

\subsection{Implications for future research}

Out of the considered origins for the radio emission (shocks, a propeller ouflow and a compact jet), a jet appears most compatible with both the correlated X-ray and radio properties and with the known properties of the Her X-1 system. The presence of a jet in Her X-1 automatically implies that (the combination of) a strong magnetic field and a slow spin do not completely impede jet formation \citep[as suggested by e.g.][]{massi08}. This would imply that our understanding of jet formation, in the presence of a strong NS magnetic field, needs to be revisited. Additionally, it opens up the possibility of observing radio emission from several currently undetected sources: for instance, the LMXBs containing slow X-ray pulsars and Be/X-ray binaries accreting from an (small) disk would be prime targets for such studies, as current generation radio telescopes (e.g. \textit{VLA}, \textit{ATCA}) reach sensitivities orders of magnitude below the current typical upper limits for these sources. 

In order to confidently confirm a jet nature of the radio emission in Her X-1, new observations are necessary. A measurement of the radio spectral index, combined with a linear polarization measurement, and a search for extended structure or a jet-break in the spectrum, could reveal the emission mechanism. If a jet is indeed present, this opens up interesting possibilities to better understand Her X-1 itself. For instance, observations at different states during its $35$-day cycle could independently confirm that precession causes this cycle, as the jet likely emits from further out without being obscured. Additionally, a jet might be a better tracer of the mass accretion rate onto the NS than the X-rays, if it is indeed not influenced by obscuration. That could possibly also allow a more detailed study of Her X-1's rare off state, wherein barely any X-ray emission is observed for extended periods of time. 

\section*{Acknowledgements}
JvdE and TDR acknowledge the hospitality of ICRAR Curtin, where part of this research was carried out, and support from the Leids Kerkhoven-Bosscha Fonds. JvdE and ND are supported by a Vidi grant from the Netherlands Organization for Scientific Research (NWO) awarded to ND. TDR is supported by a Veni grant from the NWO. JCAM-J is the recipient of an Australian Research Council Future Fellowship (FT140101082). 

%%%%%%%%%%%%%%%%%%%%%%%%%%%%%%%%%%%%%%%%%%%%%%%%%%



%\bibliographystyle{mnras}
%\bibliography{references}
\input{output.bbl}

%%%%%%%%%%%%%%%%%%%%%%%%%%%%%%%%%%%%%%%%%%%%%%%%%%

%%%%%%%%%%%%%%%%% APPENDICES %%%%%%%%%%%%%%%%%%%%%

%\appendix


%%%%%%%%%%%%%%%%%%%%%%%%%%%%%%%%%%%%%%%%%%%%%%%%%%

% Don't change these lines
\bsp	% typesetting comment
\label{lastpage}
\end{document}